\documentclass[twocolumn,aps,prl,groupedaddress,amsmath,amssymb,showpacs]{revtex4}
\usepackage{graphicx,amsmath,amssymb,epsfig}
\usepackage{dcolumn}
\usepackage{bm}
\def\pl{\partial}
\def\be{\begin{equation}}
\def\ee{\end{equation}}
\def\bea{\begin{eqnarray}}
\def\eea{\end{eqnarray}}
\def\bes{\begin{subequations}}
\def\ees{\end{subequations}}
\begin{document}
\title{Stern-Gerlach Effect of Weak-Light Ultraslow Vector Solitons}

\author{Chao Hang}
\affiliation{State Key Laboratory of Precision Spectroscopy and Department of Physics,
East China Normal University, Shanghai 200062, China}

\author{Guoxiang Huang}
\affiliation{State Key Laboratory of Precision Spectroscopy and Department of Physics,
East China Normal University, Shanghai 200062, China}

\date{\today}

\begin{abstract}

We propose a scheme to exhibit Stern-Gerlach (SG)  deflection of high-dimensional vector optical
soliton (VOS) at weak-light level in a cold atomic gas via electromagnetically induced transparency.
We show that the propagating velocity and generation power of such VOS can be reduced
to $10^{-6}\,\,c$  ($c$ is light speed in vacuum) and lowered to magnitude of nanowatt, respectively.
The stabilization of the VOS may be realized by using an optical lattice formed by a far-detuned
laser field, and its trajectory can be deflected significantly by using a SG magnetic field.
Deflection angle of the VOS can be of magnitude of $10^{-3}$ rad when propagating several millimeters.
Different from atomic SG deflection, deflection angle of the VOS can be distinct for different
polarization components and can be manipulated in a controllable way. The results obtained can
be described in terms of a SG effect for the VOS with quasispin and effective magnetic moment.

\end{abstract}

\pacs{42.65.Tg, 42.50.Gy}

\maketitle


Stern-Gerlach (SG) effect, i.e., a particle
with nonzero magnetic moment
deflects when passing through an inhomogeneous
magnetic field, was firstly discovered in early time of quantum
mechanics. This effect illustrates the necessity
for a radical departure from classical mechanics, and characterizes quantum
nature of atomic motion in a simple and fundamental way \cite{sak}. Recently, similar
effect was also predicted in many other systems, e.g., spinor Fermi
and Bose gases \cite{GO} and chiral molecules \cite{LBS}.

All massive elementary particles, such as electrons, have
non-zero magnetic moments. Contrarily, photons have no magnetic moment
in vacuum, thus experience no force when passing
through inhomogeneous magnetic field. Recently, in a very remarkable
experiment \cite{KW}, Karpa and Weitz showed that photons may acquire
effective magnetic moments when propagating in a resonant atomic gas,
and hence can deflect by a gradient magnetic field. In their experiment,
a technique of electromagnetically induced
transparency (EIT) \cite{fle} is exploited, by which a small absorption and
slow propagating velocity of photons can be realized.

However, the EIT-enhanced deflection of light in Ref. \cite{KW}
cannot be explained as a standard SG effect because
only one component of ``spin'' is involved. In this Letter, we propose
a double EIT scheme to demonstrate a SG effect of high-dimensional
vector optical soliton (VOS), which not only has two polarization
components (i.e. a quasispin) but also allows a distortionless propagation.
Propagating velocity and generation power of the VOS can be reduced
to very low level. Stabilization of the VOS can be realized using an optical lattice formed
by a far-detuned laser field. The VOS can acquire very large effective magnetic
moments, and the deflection of its trajectory is much more significant
when passing though a SG gradient magnetic field.

Before proceeding, we note that besides Ref. \cite{KW},
optical beam deflection in external fields has been the subject of
many previous works \cite{SW,hol,PUR,ZLZS,GZKS}. The present work is related to
Refs. \cite{KW,ZLZS,GZKS} and to recent studies of slow-light
solitons \cite{WD,HDP,MPP}. Essence of Refs. \cite{KW,ZLZS,GZKS} is a SG effect
of linear polaritons. However, such linear polaritons spread
and attenuate during propagation because of the existence of diffraction
and other detrimental effects. In Refs. \cite{WD,HDP,MPP},
slow-light solitons via EIT are suggested, but no SG effect
is considered.

In contrast, the scheme presented here exploits optical lattice
and EIT-enhanced Kerr effect, which allow the formation and stable propagation of
high-dimensional VOS, or called {\it nonlinear polariton}, with effective magnetic moment
(SG deflection)  being   four (two) orders of magnitude larger
than that of the linear polariton of Ref. \cite{KW}. Thus, comparing with that
obtained in a linear scheme \cite{KW,ZLZS,GZKS}, the SG effect proposed
here is more efficient and robust for observation
and practical applications.

\begin{figure}
\centering
\includegraphics[width=8cm]{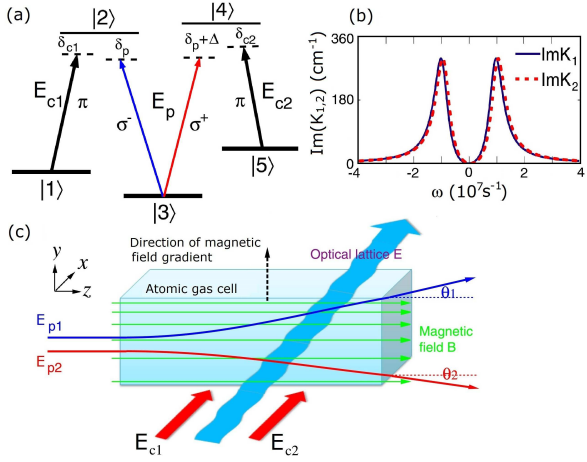}
\caption{\footnotesize{(color online) (a): Double EIT scheme. ${\bf E}_p$ and
 ${\bf E}_{cj}$ ($j=1,2$) are probe and control fields, respectively; $\delta_p$, $\delta_p+\Delta$,
 and $\delta_{cj}$ are detunings. (b): Absorption spectrum Im$K_j(\omega)$ as functions
of $\omega$. Solid and dotted lines
correspond to the $\sigma^-$  and $\sigma^+$
polarization components, respectively.
(c): A possible experimental arrangement, where an
inhomogeneous magnetic field ${\bf B}$
removes the degeneracy of ground states $|j\rangle$ ($j=1,3,5$)
and excited states  $|l\rangle$ ($l=2, 4$), and
causes Stern-Gerlach deflection of probe-field components. $\theta_1$ and $\theta_2$ are deflection
angles of $\sigma^{-}$  polarization component (i.e. ${\bf E}_{p1}$) and $\sigma^{+}$ polarization component
(i.e. ${\bf E}_{p2}$) of high-dimensional VOS, which has a quasispin and an effective
magnetic moment. The curved thick arrow represents the far-detuned optical lattice
field ${\bf E}$ used to stabilize the VOS.
}}\label{Fig1}
\end{figure}

To be specific, we consider a medium consisting of five-level atoms
with M-configuration. A linearly polarized,
pulsed probe field (with pulse duration $\tau_0$) ${\bf E}_{p}={\bf E}_{p1}+{\bf E}_{p2}=({\bf
\hat{\epsilon}}_{-}{\cal E}_{p1} + {\bf \hat{\epsilon}}_{+}{\cal
E}_{p2}) \exp [i (k_p z-\omega_p t)]+{\rm c.c.}$ drives the transitions
$|3\rangle\leftrightarrow |2\rangle$ and $|3\rangle\leftrightarrow|4\rangle$
by its left-circular (i.e. $\sigma^{-}$) polarization component ${\bf E}_{p1}$ and
right-circular (i.e. $\sigma^{+}$) polarization component ${\bf E}_{p2}$,
respectively. Here ${\cal E}_{p1,p2}$ are envelopes and
${\bf \hat{\epsilon}}_{\pm } \equiv ({\bf \hat{x}}\pm i{\bf \hat{y}})/\sqrt{2}$.
A $\pi$-polarized, strong continuous-wave control field ${\bf E}_{c1}={\bf \hat{z}} {\cal
E}_{c1} \exp [i (k_{c1} x-\omega_{c1} t)]+{\rm c.c.}$
(${\bf E}_{c2}={\bf \hat{z}} {\cal E}_{c2} \exp [i (k_{c2} x-\omega_{c2} t)]
+{\rm c.c.}$) drives the transition $|1\rangle\leftrightarrow
|2\rangle$ ($|5\rangle\leftrightarrow |4\rangle$) (Fig. \ref{Fig1}(a)\,).
$\hat{\bf x}$, $\hat{\bf y}$, and $\hat{\bf z}$
are unit vectors along coordinate axes $x$, $y$ and $z$, respectively  (Fig. \ref{Fig1}(b)).

%
%

We assume an inhomogeneous magnetic field
${\bf B}(y)=\hat{\bf z}B(y)=\hat{\bf z} (B_0+B_{1}y)$ ($B_1\ll B_0$)
is applied to the system. Here $B_0$ contributes to a Zeeman level shift
$\Delta E_{\rm Z}=\mu_B g_F^{j} m_F^{j} B_0$, and hence
removes the degeneracy of ground-state sublevels $|j\rangle$ ($j=1,3,5$)
and the excited-state sublevels $|l\rangle$ ($l=2, 4$). $\mu_B$, $g_F^{j}$,
and $m_F^{j}$ are Bohr magneton, gyromagnetic factor, and magnetic quantum
number of the level $|j\rangle$, respectively. $B_1$ contributes
a transverse gradient of the magnetic field,
resulting in a SG deflection of the probe field.

We assume further a small, far-detuned laser field
${\bf
E}(x,t)=\hat{\bf x}E_{0} \cos (x/R_{\perp}) \cos(\omega_L t)$
is also applied into the medium, where $E_0$, $R_{\perp}$, and
$\omega_L$ are field amplitude, beam radius, and angular
frequency, respectively. Due to ${\bf E}(x,t)$, Stark level shift
$\Delta E_{j,\rm S} =-\frac{1}{2}\alpha_j \langle
E^2\rangle_t=-\frac{1}{2} \alpha_j E^2(x)$ occurs, here $\alpha_j$
is the scalar polarizability of the level $|j\rangle$, $\langle
\cdots \rangle_t$ denotes the time average in an oscillation cycle, and
hence $E(x)=(E_0/\sqrt{2}) \cos (x/R_{\perp})$.
The aim of introducing the far-detuned laser field
is to form an optical lattice potential to stabilize the
high-dimensional VOS \cite{BMS}, as shown below.

Besides, atoms are assumed prepared initially in the ground-state level
$|3\rangle$ and trapped in a gas cell with ultracold temperature to cancel
Doppler broadening and collisions.
Thus, the system is composed of two $\Lambda$-type EIT configurations
(i.e. double EIT). A possible arrangement of experimental
apparatus is suggested in Fig. \ref{Fig1}(c).

Under electric-dipole and rotating-wave approximations, the Hamiltonian of
the system in interaction picture is
$H_{\rm int}/\hbar=(\delta_p-\delta_{c1})|1\rangle\langle
1|+\delta_p|2\rangle\langle 2|+(\delta_p+\Delta)|4\rangle\langle
4|+(\delta_p+\Delta-\delta_{c2})|5\rangle\langle
5|+\Omega_{c1}|2\rangle\langle1|+\Omega_{p1}|2\rangle\langle3|+\Omega_{p2}|4\rangle\langle3|
+\Omega_{c2}|4\rangle\langle5|+{\rm H.c.}$,
where $\Omega_{p1}$=$-(\textbf{p}_{23}\cdot \hat{{\bf \epsilon}}_-){\cal E}_{p1} /\hbar$ and
$\Omega_{p2}$=$-(\textbf{p}_{43}\cdot \hat{{\bf \epsilon}}_+){\cal E}_{p2}/\hbar$
($\Omega_{c1}$=$-(\textbf{p}_{21}\cdot \hat{\bf z}){\cal E}_{c1}/\hbar$ and
$\Omega_{c2}$=$-(\textbf{p}_{45}\cdot \hat{\bf z}){\cal E}_{c2} /\hbar$) are respectively
Rabi frequencies of two circularly polarized components of the probe
field (two $\pi$-polarized control fields), with
$\textbf{p}_{jl}$ being the electric dipole matrix element
associated with the transition from $|j\rangle$ to $|l\rangle$.
The detunings are defined as
$\delta_p=\omega_{23}+\mu_{23}B(y)-\frac{1}{2}\alpha_{23} E(x)^2-\omega_p$,
$\delta_{c1}=\omega_{21}+\mu_{21}B(y)-\frac{1}{2}\alpha_{21} E(x)^2-\omega_{c1}$,
$\delta_{c2}=\omega_{45}+\mu_{45}B(y)-\frac{1}{2}\alpha_{45} E(x)^2-\omega_{c2}$,
and $\Delta=\mu_{42}B(y)-\frac{1}{2}\alpha_{42} E(x)^2$, where
$\mu_{jl}=\mu_B(g_F^{j}m_F^{j}-g_F^{l}m_F^{l})/\hbar$,
$\alpha_{jl}=(\alpha_{j}-\alpha_{l})/\hbar$,
and $\omega_{jl}=(E_j-E_l)/\hbar$, with $E_j$ being the eigenenergy of the state $|j\rangle$.

The motion of atoms is governed by the Bloch equation for
density-matrix $\rho$,
\be
\partial \rho/\partial t=-i[H_{\rm
int},\rho]/\hbar-\Gamma(\rho),
\ee
where $\Gamma(\rho)$ is relaxation matrix representing spontaneous
emission and dephasing (see Supplementary Material).
Electric-field evolution is controlled by Maxwell equation
\be\label{Maxwell}
\nabla^{2}\textbf{E}-(1/c^{2})\pl^{2}\textbf{E}/\pl t^{2}
=(1/\epsilon_{0}c^{2})\pl^{2}\textbf{P}/\pl t^{2},
\ee
where
$\textbf{P}={\cal N}{\rm Tr}({\bf p}\rho)$ is electric
polarization with ${\cal N}$ the atomic density. Under slowly
varying envelope approximation, Eq. (\ref{Maxwell})  reduces to
$[i\partial_z+(i/c)\partial_t+c\nabla_{\perp}^2/(2\omega_p)]\Omega_{p1,p2}
-\kappa_{32,34} \rho_{23,43}=0$,
where $\nabla_{\perp}^2=\partial^2/\partial x^2+\partial^2/\partial y^2$ and
$\kappa_{32,34}={\cal N} |\textbf{p}_{32,34}\cdot{\bf
\hat{\epsilon}}_{\mp}|^2\omega_p/(2\hbar\epsilon_0c)$ with $\epsilon_{0}$ the vacuum dielectric
constant.


Linear propagation of
the probe field in the absence of diffraction can be obtained by taking $\Omega_{p1,p2}$ as
small quantities and $B_1$, $E_0$ as zero. Then one has
$\Omega_{pj}=F_j \exp \{i[K_j(\omega)z-\omega t)]\}$ ($j=1,2$)
with
$K_{1,2}(\omega)=\omega/c+\kappa_{32,34}(\omega-d_{1,5})/D_{1,2}$
(linear dispersion relation).
Here, $F_{j}$ are constants,
$D_{1,2}=|\Omega_{c1,c2}|^{2}-(\omega-d_{1,5})(\omega-d_{2,4})$,
$d_{1}=(\delta_p-\delta_{c1})-i\gamma_{13}/2$,
$d_{2}=\delta_p-i(\Gamma_2+\gamma_{23})/2$,
$d_{4}=(\delta_p+\Delta)-i(\Gamma_4+\gamma_{34})/2$, and
$d_{5}=(\delta_p+\Delta-\delta_{c2})-i\gamma_{35}/2$ with
$\delta_p=\omega_{23}+\mu_{23}B_0-\omega_p$, $\Delta=\mu_{42}B_0$,
$\delta_{c1}=\omega_{21}+\mu_{21}B_0-\omega_{c1}$, and
$\delta_{c2}=\omega_{45}+\mu_{45}B_0-\omega_{c2}$. $\Gamma_{i}$
and $\gamma_{ij}$ denote the spontaneous emission and dephasing
rates of relevant states, respectively.

The linear dispersion relation displays two branches.
Fig. \ref{Fig1}(b) shows the absorption spectrum
of Im$K_j(\omega)$ ($j=1,2$) as a function of
frequency $\omega$. Parameters are chosen for a laser-cooled $^{85}$Rb atomic gas
with $|1\rangle=|5
^2S_{1/2},F=2,m_F=-1\rangle$, $|2\rangle=|5
^2P_{1/2},F=2,m_F=-1\rangle$,
$|3\rangle=|5^2S_{1/2},F=1,m_F=0\rangle$,
$|4\rangle=|5^2P_{1/2},F=2,m_F=1\rangle$, and
$|5\rangle=|5^2S_{1/2},F=2,m_F=1\rangle$. Decay rates are
$\Gamma_2\simeq\Gamma_4\simeq6$ MHz and
$\gamma_{13}\simeq\gamma_{23}\simeq\gamma_{34}\simeq\gamma_{35}\simeq50$
Hz. Other parameters are taken as
$\kappa_{32}\simeq\kappa_{34}=1.0\times10^9$ cm$^{-1}$s$^{-1}$,
$\Omega_{c1}=\Omega_{c2}=1.0\times10^{7}$ s$^{-1}$,
$\delta_p=\delta_{c1}=\delta_{c2}=0$, and $B_0=34.1$ mG. The solid (dotted) line
in the figure is for $\sigma^{-}$ ($\sigma^{+}$) polarization component.
We see that large and deep transparency windows in the
absorption spectra of both polarization components (double EIT) appear.
Using above parameters, group velocities of the both components (defined by $V_{gj}={\rm
Re}(\partial K_j/\partial \omega)^{-1}$) are given as
$3.3\times10^{-6}c$.


However, the linear solution is unstable due to the
diffraction and other detrimental effects, which results in spreading and attenuation
of the probe field during propagation, as demonstrated by Eq. (24)
of Ref. \cite{GZKS}. To solve this problem we
use nonlinear effect to suppress the spreading and attenuation. When including weak nonlinearity and
diffraction, we obtain the following nonlinearly coupled,
dimensionless equations, derived by using a standard method of multiple-scales
(see the Supplementary Material):
\bea\label{NLS1}
& & \left[\frac{i}{v_{g1,g2}}\frac{\partial }{\partial \tau}
+\frac{1}{2}\left(\frac{\partial^{2} }{\partial
\xi^{2}}+\frac{\partial^{2} }{\partial
\eta^{2}}\right)\right]u_{1,2}-\frac{1}{\sqrt{2\pi}\rho_0}(g_{11,22}|u_{1,2}|^{2}\nonumber\\
& &
+g_{12,21}|u_{2,1}|^{2})u_{1,2}+V_{1,2}(\xi,\eta)u_{1,2}=-iA_{1,2}u_{1,2}.
\eea
where $\tau$=$t/\tau_0$,
$(\xi,\eta)$=$(x,y)/R_{\bot}$, $v_{gj}$=$V_{gj}\tau_0/L_{\rm Diff}$, and $u_{j}(\tau,\xi,\eta)$=$[\Omega_{pj}/F_j(\rho_j)]
e^{-i{\rm Re}[K_j|_{\omega=0}]z}/U_0$. Here, $L_{\rm Diff}=\omega_p R_{\bot}^2/c$ and $U_0$ are
respectively the typical diffraction length and Rabi frequency,
$F_{j}(\rho_j)$ are normalized Gaussian functions (i.e.
$F_{j}=[1/(\rho_0\sqrt{\pi})]^{1/2}\exp [-\rho_j^2/(2\rho_0^2)]$
with $\rho_j=(z-V_{gj}t)/L_{\rm Diff}$ and $\rho_0$ a constant \cite{GZKS}),
$g_{11,12,21,22}=W_{11,12,21,22}/|W_{22}|$ are nonlinearity coefficients with $W_{11,22}=
-\kappa_{32,34}d_{1,5}(|d_{1,5}|^{2}+|\Omega_{c1,c2}|^{2})/(D_{1,2}|D_{1,2}|^{2})$ and
$W_{12,21}=-\kappa_{32,34}d_{1,5}(|d_{5,1}|^{2}+|\Omega_{c2,c1}|^{2})/(D_{1,2}|D_{2,1}|^{2})$
characterizing respectively self-phase and cross-phase modulations, and
$A_{j}$=${\rm Im}[K_j|_{\omega=0}]L_{\rm Diff}$ are small absorption coefficients.

Combined potentials in Eq. (\ref{NLS1}) have the form
\be \label{potential}
V_{j}(\xi,\eta)={\cal M}_{j}\eta+{\cal N}_{j}\cos^2(\xi),
\ee
where
${\cal M}_{1,2}=M_{1,2}R_{\bot}B_{1}$  and ${\cal N}_{1,2}=N_{1,2}E_{0}^2$
are contributions from the SG gradient magnetic field (proportional to $B_1$)
and the optical lattice field (proportional to $E_0^2$),
respectively. $M_{j}$ and $N_{j}$ are defined as
$M_{1,2}$= $-\kappa_{32,34}(d_{1,5}^{2}\mu_{23,43}+|\Omega_{c1,c2}|^2\mu_{13,53}) L_{\rm
Diff}/D_{1,2}^2$ and $N_{1,2}$=$\frac{1}{4}\kappa_{32,34}(d_{1,5}^{2}\alpha_{23,43}
+|\Omega_{c1,c2}|^2\alpha_{13,53}) L_{\rm Diff}/D_{1,2}^2$.
When deriving Eq. (\ref{NLS1}), $B_1$ and $E_0^2$ are assumed as small quantities.
Additionally, $\tau_0$ is also assumed to be large (e.g. $\tau_0=2.1$ $\mu$s) so that
the second-order dispersion (proportional to $\partial^2 u_j/\partial \tau^2$)
is negligible.

\begin{figure}
\centering
\includegraphics[width=7cm]{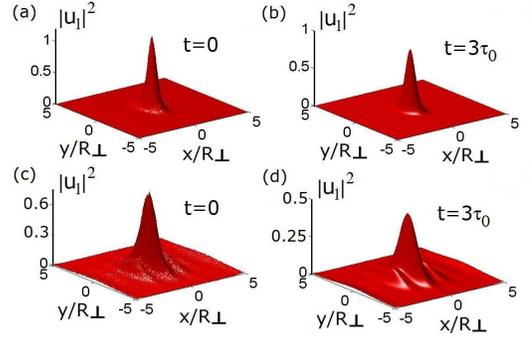}
\caption{ \footnotesize{ (color online) (a) and (b):
Evolutions of $|u_1|^2$ respectively at $t=0$ and $t=3\tau_0$
for single-peaked VOS. (c) and (d): Evolutions of $|u_1|^2$ respectively
at $t=0$ and $t=3\tau_0$ for multiple-peaked VOS.
SG gradient magnetic field is absent (i.e. $B_1=0$).
The stability of the VOS is achieved by the far-detuned optical lattice.
Result for $|u_2|^2$ is similar to $|u_1|^2$ thus
not shown.} } \label{Fig2}
\end{figure}

Fig. \ref{Fig2} (a) and (b) show
%
%
 results of numerical simulation for $|u_1|^2$ respectively at $t=0$ and $t=3\tau_0$
 for a deep optical lattice ($E_0=3.2\times 10^4$ V cm$^{-1}$). The soliton obtained
 displays a single-peaked structure. The result for $|u_2|^2$ is similar to $|u_{1}|^2$
 due to symmetry and hence not shown. The case for a shallower optical lattice
 ($E_0=2.3\times 10^4$ V cm$^{-1}$) is also simulated, with the result plotted
 in panels (c) and (d) for $t=0$ and $t=3\tau_0$,
respectively. We see that in this case a multiple-peaked soliton appears.
In both simulations, $\delta_p=1.0\times10^{6}$
s$^{-1}$, $\delta_{c2}=1.0\times10^{5}$ s$^{-1}$, and $R_{\perp}=16$ $\mu$m with other parameters
the same with those in Fig. \ref{Fig1}. In addition, $U_0=6.8\times 10^6$
s$^{-1}$, which allows enough nonlinearity to balance the
diffraction. The typical diffraction length $L_{\rm Diff}$
and nonlinearity length $L_{\rm Nonl}$  ($\equiv 1/(U_0^2|W_{22}|$) are $\simeq0.2$ cm.
Furthermore, $B_1$ is chosen as zero, i.e., the SG
gradient magnetic field is absent, thus no SG deflection occurs.

The stability of the high-dimensional
VOS is checked by adding a small random perturbation to the
stationary solution obtained in imaginary time (Fig. \ref{Fig2} (a),
(c)) and evolving the solution according to Eq. (\ref{NLS1}) in real
time. We find that the soliton can indeed propagate stably for a
long time (Fig. \ref{Fig2} (b), (d)). We have also used a
standard linear stability analysis (see the Supplementary Material) to confirm
the stability of the high-dimensional
VOS.

\begin{figure}
\centering
\includegraphics[width=8cm]{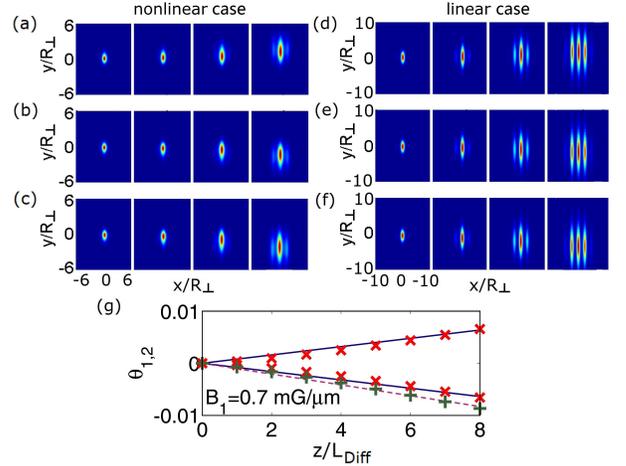}
\caption{ \footnotesize{(color online) SG effect of ultraslow VOS. (a) and (b): Symmetric deflection
(on $y$-axis) of $|u_1|^2$ and $|u_2|^2$ when propagating from $z=2
L_{\rm Diff}$ to $z=8 L_{\rm Diff}$ (corresponding respectively to the
subfigure from left to right), respectively.
(c): Asymmetric deflection of $|u_2|^2$
($|u_1|^2$ is the same as (a) thus not shown). (d), (e), (f):
Corresponding evolution of linear polariton.
(g):  Deflection angles of the VOS
as functions of $z/L_{\rm Diff}$ for $B_1=0.7$ mG/$\mu$m.
The solid line with positive (negative) slope is the analytical result of $\theta_1$ ($\theta_2$) for the
symmetric case. Dashed line is the analytical result of $\theta_2$ for the
asymmetric case ($\theta_1$ is the same as the symmetric case thus not shown).
Points labeled by ``x'' and ``+'' are center positions of the VOS polarization components obtained
numerically. \label{Fig3}} }
\end{figure}

Next we study VOS deflection by
numerically simulating Eq. (\ref{NLS1}) with $B_1\neq0$. Shown in Fig. \ref{Fig3}
%
%
(a) and (b) are spatial distributions of $|u_1|^2$ (panel (a)) and
$|u_2|^2$ (panel (b)) in ($x,y$)-plane when the VOS propagates from $z=2
L_{\rm Diff}$ to $z=8 L_{\rm Diff}$ with group velocity
$V_{g1}\simeq V_{g2}=3.2\times10^{-6}c$. In the simulation,
$B_1=0.7$ mG $\mu$m$^{-1}$ is chosen. We see that an obvious
deflection of VOS trajectories occurs due to the existence of the
SG gradient magnetic field. Additionally, two different
polarization components deflect symmetrically in
$+y$ and $-y$ directions, similar to the SG deflection for atoms.

The SG deflection of VOS components can be made asymmetric. To show this,
we take $\Omega_{c2}=0.9\times10^{7}$ s$^{-1}$ without changing
other parameters, then $(V_{g1},V_{g2})=(3.2,2.6)
\times10^{-6}c$. As a result, the trajectory of $\sigma^-$ component
keeps unchanged, whereas the trajectory of $\sigma^+$
component changes as shown in Fig. \ref{Fig3}(c).
This is different from atomic SG deflection, where trajectories are
always symmetric for two different spin components.

For comparison, in Fig. \ref{Fig3}(d), (e), and (f) we present results of corresponding evolution
for a linear polariton. One sees that the probe pulse spread rapidly.
Thus the nonlinear effect is necessary for obtaining stable VOS and its robust SG deflection.

Analytical VOS solutions of Eq. (\ref{NLS1}) can be gained under some approximations:
(i)The small
absorption term $-iA_{j}u_{j}$ is disregarded. (ii)Since in
the presence of the SG gradient magnetic field the two
polarization components of VOS separate each other after
propagating some distance, the cross-phase-modulation terms can be
neglected. (iii)The optical lattice is deep enough so that $V_j$ can be
approximated as ${\cal M}_j\eta+{\cal N}_j(1-\xi^2)$. Taking
$u_j(\tau,\xi,\eta) =w_j(\tau,\eta) \phi_j(\xi) \exp[i{\cal
N}_jv_{gj}\tau]$, where $\phi_j(\xi)$ is the normalized ground
state of the eigenvalue problem $(\partial^2/\partial \xi^2-2{\cal
N}_{j}\xi^2)\phi_j=2E_{\xi}\phi_j$ with $E_{\xi}=-\sqrt{{\cal
N}_{j}/2}$, and integrating out the variable $\xi$, Eq.
(\ref{NLS1}) reduces to
$[(i/v_{gj})\partial_\tau+(1/2)\partial^{2}_\eta]w_{j}-{\cal N}_j^{1/4}/(2^{3/4}
\pi\rho_0)g_{jj}|w_{j}|^{2}w_{j}+({\cal M}_j\eta-\sqrt{{\cal N}_j/2})w_{j}=0$,
which admits exact soliton solutions \cite{YZY}. A single-soliton
solution (see the Supplementary Material) gives
\bea \label{Sol}
\Omega_{pj}=& & U_0 A_j\,[1/(\rho_0\sqrt{\pi})]^{1/2}\,(\sqrt{2{\cal N}_j}/\pi)^{1/4}
e^{i\varphi_j} \nonumber\\
& & \times e^{-(s-v_{gj}\tau)^2/(2\rho_0^2)}\, e^{-\sqrt{{\cal N}_j}\xi^2/\sqrt{2}}{\rm
sech}\Theta_j,
\eea
where  $A_j=(2^{5/4}{\cal N}_j^{1/4}\pi\rho_0/|g_{jj}|)^{1/2}$, $\varphi_j={\cal
M}_jv_{gj}\tau(\eta-{\cal M}_jv_{gj}^2\tau^2/6)$, and
$\Theta_j=(2{\cal N}_j)^{1/4}(\eta-{\cal
M}_jv_{gj}^2\tau^2/2)$ ($j=1,2$).
We see that both VOS components are localized in three spatial and one temporal dimensions.
Thus, $(u_1, u_2)$ can be considered as a {\it vector light bullet}.

After passing the medium with length $L$, the center position of the $j$th polarization
component of the VOS is at $(x,y_j,z)$=$(0,{\cal M}_jL^2 R_{\bot}/(2L_{\rm diff}^2),L)$, with
the propagating velocity along the $z$- ($y$-) direction given by $V_{gj}$ ($V_{j}\equiv {\cal
M}_jv_{gj}^2R_{\bot}t/\tau_0^2)$. As a result, the expected deflection angle of the $j$th
VOS component is
\be\label{Angle}
\theta_j=V_{j}/V_{gj}=(L/V_{gj})(\mu_{{\rm sol}\,j}/p)r^2B_{1},
\ee
where  $r=R_{\bot}/L_{\rm Diff}$, $p=\hbar k_p$ is photon momentum,
$\mu_{{\rm sol}\,j}=M_{j} V_{gj}\hbar k_p$ is effective magnetic moment.
With the data in Fig. \ref{Fig3}, we obtain $\mu_{{\rm sol}\,1,2}=\pm7.6\times10^{-20}$ J/T,
which is four orders of magnitude larger than the effective magnetic moment for
linear polariton of Ref. \cite{KW}. From Eq. (\ref{Angle})
we see the deflection angle of the $j$th polarization component of the VOS
is proportional to the medium length $L$, the SG gradient magnetic field $B_{1}$,
and inversely proportional to the group velocity $V_{gj}$. In a mechanical viewpoint, the
deflection of the $j$th component of the VOS is caused by the
transverse magnetic force $F_j=\mu_{{\rm sol}\,j}B_{1}$ and deflection angles
can be expressed as $\theta_j=F_jt_{{\rm int}\,j}r^2/p_j$
with $t_{{\rm int}\,j}=L/V_{gj}$ being the interaction time.
Due to untraslow propagating velocity,
large deflection angles may be observed even for small $L$.

Fig. \ref{Fig3}(g) shows deflection angles of the VOS
as functions of $z/L_{\rm Diff}$ for $B_1=0.7$ mG/$\mu$m.
The solid line of positive (negative) slope is the analytical result
of $\theta_1$ ($\theta_2$)  by Eq. (\ref{Angle}) for
the $\sigma^-$ ($\sigma^+$) component with $V_{g1}\simeq V_{g2}=3.3\times10^{-6}c$ (i.e.
the symmetric case). Points labeled by ``x'' are center positions of the VOS
components obtained numerically. We thus have
$(\theta_1,\theta_2) \simeq (1.6,-1.6) \times 10^{-3}$ rad for $z=0.4$ cm,
which is two orders of magnitude larger than that for
linear polariton obtained in Ref. \cite{KW}.
The dashed line is the analytical result of $\theta_2$ with
$(V_{g1},V_{g2})=(3.2,2.6)\times 10^{-6}c$ (i.e. the asymmetric case) and points
labeled by ``+'' are numerical results
($\theta_1$ is the same as the symmetric case).
In both cases analytical results agree well with
numerical ones.

The generation power of the
high-dimensional VOS predicted above can be estimated by using Poynting's vector \cite{HDP},
which is $3.5$ nW calculated using the above parameters. Thus, very low input power is needed
for generating the VOS in the present double EIT system.

In conclusion,  a scheme is proposed to exhibit SG deflection of high-dimensional
VOS via a double EIT. The VOS has ultraslow propagating velocity
and extremely low generation power.
The stabilization of the VOS can be realized by using an optical lattice,
and its trajectory can be significantly deflected  by a
SG gradient magnetic field.  The results obtained can be described in terms
of a SG effect of the VOS with quasispin and effective magnetic moments.
We expect that such large and  robust SG effect may have potential applications in
magnetometery and quantum information processing.

This work was supported by the NSF-China
under Grant Nos. 11174080 and 11105052.




\begin{references}

\bibitem{sak} J. J. Sakurai, {\it Modern Quantum Mechanics} (Revised Edition) (Addison-Wesley, 1994).

\bibitem{GO} M. D. Girardeau and M. Olshanii, Phys. Rev. A {\bf 70}, 023608 (2004).

\bibitem{LBS} Y. Li {\it et al.}, Phys. Rev. Lett. {\bf 99}, 130403
(2007).

\bibitem{KW} L. Karpa and M. Weitz, Nat. Phys. {\bf 2}, 332 (2006).

\bibitem{fle} M. Fleischhauer {\it et al}., Rev. Mod. Phys. {\bf 77}, 633 (2005).

\bibitem{SW} R. Schlesser and A. Weis, Opt. Lett. {\bf 17}, 1015 (1992).

\bibitem{hol} R. Holzner {\it et al.}, Phys. Rev. Lett. {\bf 78}, 3451 (1997).

\bibitem{PUR}G. T. Purves {\it et al}., Eur. Phys. J. D {\bf 29}, 433 (2004).

\bibitem{ZLZS} D. L. Zhou {\it et al.}, Phys. Rev. A {\bf 76}, 055801 (2007).

\bibitem{GZKS} Y. Guo {\it et al.}, Phys. Rev. A {\bf 78}, 013833 (2008).

\bibitem{WD} Y. Wu and L. Deng, Phys. Rev. Lett. {\bf 93}, 143904 (2004).

\bibitem{HDP} G. Huang {\it et al.}, Phys. Rev. E {\bf 72},
             016617 (2005).

\bibitem{MPP} H. Michinel {\it et al.},
              Phys. Rev. Lett. {\bf 96}, 023903 (2006).

\bibitem{BMS} B. B. Baizakov {\it et al.}, Phys. Rev. A {\bf 70}, 053613 (2004).




\bibitem{YZY} J. Yan {\it et al.}, Phys. Fluids  A {\bf 4}, 690 (1992).








\end{references}
\end{document}